\documentclass[10pt,letterpaper,twocolumn]{article} %% two column, final layout

\usepackage{ol}
\usepackage{hyperref}
\usepackage{amsmath}

\begin{document}

\twocolumn[ %% activate for two-column option

\title{Quadratic Phase Matching in slot waveguides}

\author{Andrea Di Falco}
\address{
N\it{oo}EL - Nonlinear Optics and OptoElectronics Lab, University ``Roma Tre''- Rome, Italy }

\author{Claudio Conti}
\address{
Research Center ``Enrico Fermi'', Via Panisperna 89/A - Rome, Italy\\
Research Center Soft INFM-CNR, University ``La Sapienza'', Piazzale A. Moro, 2 - 00185 - Rome, Italy}

\author{Gaetano Assanto}
\address{
N\it{oo}EL - Nonlinear Optics and OptoElectronics Lab \\
Department of Electronic Engineering, INFN and CNISM, University ``Roma Tre''\\
Via della Vasca Navale, 84 - 00146 - Rome, Italy }

\begin{abstract}
We analyze phase matching with reference to frequency doubling in nanosized quadratic waveguides encompassing form birefringence and supporting cross-polarized fundamental and second harmonic modes.
In an AlGaAs rod with an air-void, we show that phase-matched second-harmonic generation could be achieved in a wide spectral range employing state-of-the-art nanotechnology.
\end{abstract} 


 ] %% activate for two-column option

\noindent Since the pioneering work by Franken {\it et al.}, \cite{Franken} frequency doubling has been among the most 
studied nonlinear processes in optics. Conversion efficiency in second 
harmonic generation (SHG) relates to spatio-temporal overlap of  
waves at fundamental (FF) and second harmonic (SH) frequencies, as well as to their 
group and phase velocity matching. For the latter, in order to compensate material and waveguide 
dispersion, several approaches have been proposed and demonstrated, \cite{Armstrong} 
including birefringent phase-matching (PM), non-critical PM, 
quasi-PM (QPM) \cite{Boyd} in co- or counter-propagating geometries, 
including vertical QPM for surface-emitted SHG. \cite{Fejer,Ding,Vella_APL1981,RavaroJAP_05} 
More recently, novel PM schemes for SHG include random quasi-PM, \cite{Baudrier} 
photonic crystals and microcavities, \cite{Levenson,DiFalco} as well as 
form-birefringence in waveguides.\cite{VanderZiel,Fiore,Moutzouris,Berger_JQE}   
Form-birefringence conjugates the large intensities and interaction distances of guided-waves  
with the possibility of tailoring the dispersion of cross-polarized waves even in isotropic crystals 
such as GaAs and its composites.\cite{Fiore,Moutzouris,Berger_JQE}    
It consists in alternating layers with different refractive indices orthogonally to the direction 
of propagation, such that the harmonics experience distinct 
eigen-values for TE (quasi-TE) and TM (quasi-TM) polarizations. As in photonic crystals, the degree of birefringence, its dispersion and the achievable PM line-width are 
directly linked to the index contrast between alternating films and their thicknesses,\cite{Joannopoulos} making film growth a formidable task.\cite{Berger_JQE}  For these reasons, a nano-sized structure encompassing form-birefringence but realizable with just one nonlinear 
material should provide significant advantages and versatility over standard approaches for parametric mixing and frequency generation.

\begin{figure}[htb]
\centerline{\includegraphics[width=8.3cm]{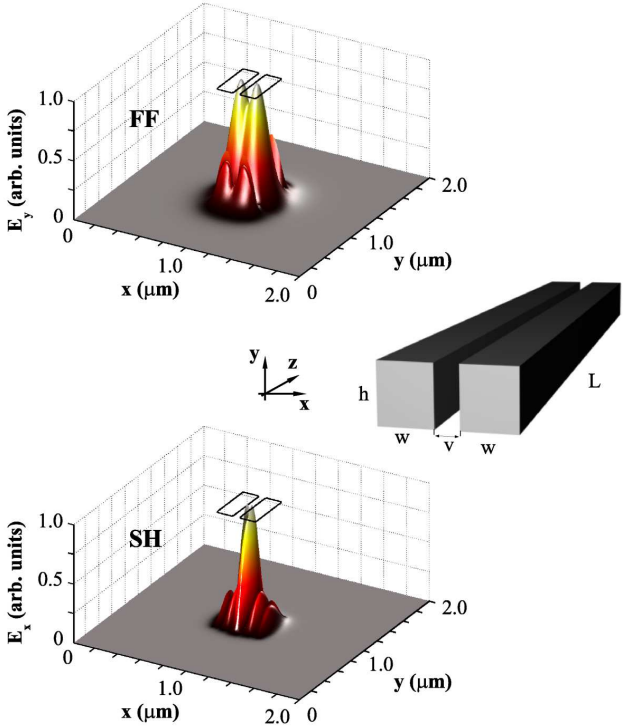}}
 \caption{Sketch of the air-void waveguide and corresponding lowest-order 3D eigen-profiles. Top: quasi-TM FF; bottom: quasi-TE SH. 
Here $h=400nm$, $w=150nm$, $v=50nm$ and $\lambda=1550nm$ at FF.}
\label{fig0}
\end{figure}

In this Letter we propose and investigate SHG in a suspended rod-waveguide with a nanometric void, 
implementing form-birefringence and large modal overlap for efficient upconversion. 
To achieve the highest index contrast between a quadratically nonlinear semiconductor (AlGaAs) 
and air (or vacuum), we consider the basic geometry sketched in Fig. \ref{fig0}, with the electric 
field of the FF mode parallel to the void (i.e., quasi-TM with respect to the top surface) and a cross-polarized (i.e., quasi-TE) SH 
tightly confined in the air-gap owing to the discontinuity in the electric field normal to the 
interfaces, the latter separated by $v$. Fig. \ref{fig0} also displays the calculated FF and SH lowest-order eigen-profiles 
of such a slot waveguide, similar to the one recently studied by Xu {\it et al.} for ring-resonators in $SiO_2$.\cite{Lipson}
We consider cubic $Al_{0.3}Ga_{0.7}As$ with indices $n_{FF}$ and $n_{SH}$ at FF and SH, 
respectively, after Ref. [\citeonline{Afromovitz}]. The insertion of a symmetrically-placed air-gap of width $v$ allows to phase-match the 
quadratic interaction between a quasi-$TM_{00}$ mode of wavelength $\lambda_{FF}$ and its quasi-$TE_{00}$ second harmonic, 
leaving nearly unaltered the effective second-order susceptibility.

To explore PM in the structure, we fixed the slot-waveguide height ($h=400 nm$, see Fig. \ref{fig0}) as to allow single-mode confinement 
at FF and form-birefringence PM for SHG over the explored spectrum. Variations in $h$ modify the pertinent propagation constants and the PM wavelengths, but only marginally affect the modal overlap integrals and the conversion.

%FF and SH profiles are modified accordingly, hence the overall analysis is performed focusing on the other parameters.

We calculated the effective indices versus slot aperture $v$ and individual rod-width $w$ resorting 
to a finite-difference semi-vectorial mode-solver. This was validated against a (computationally heavy) fully-vectorial solver and yielded accurate results for the nanometric geometry under investigation.  We also exploited symmetry and solved the eigenvalue problem for a quarter of the (transverse) geometry.

\begin{figure}[htb]
\centerline{\includegraphics[width=8.3cm]{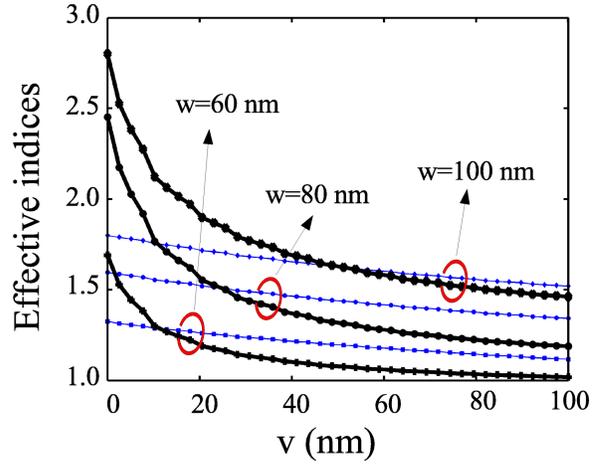}}
 \caption{Effective indices at FF (thick lines) and SH (thin lines) vs void $v$ for $\lambda_{FF}=1550 nm$ and three values of $w$. Crossings of paired curves correspond to type I PM for SHG.}
\label{fig1}
\end{figure}

Fig. \ref{fig1} shows the computed effective indices $\beta\lambda/2\pi$ for FF (thick lines) and SH (thin lines) versus $v$ and for three widths $w$ at $\lambda_{FF}=1550 nm$. The curves at $\lambda_{SH}=\lambda_{FF}/2$ exhibit a much stronger dispersion due to the filling-factor of the eigensolution in the gap: 
the SH mode is much more affected by small changes in $v$ than the FF, the latter exhibiting a nearly linear dispersion. 
Clearly, this becomes more appreciable as $v$ approaches zero, whereas dispersion flattens for large $v$. 

\begin{figure}[htb]
\centerline{\includegraphics[width=8.3cm]{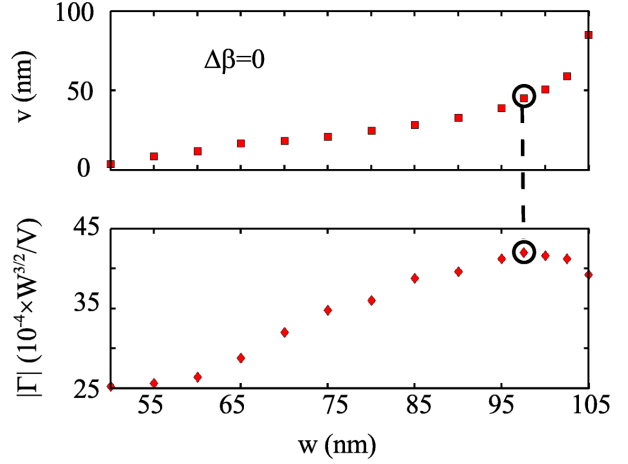}}
 \caption{Top: PM condition ($\Delta\beta=0$) versus $v$ and $w$ (in $nm$); bottom: corresponding moduli of the SHG overlap integral. Circles indicate the values providing maximum conversion efficiency.}
\label{fig2}
\end{figure}

Fig. \ref{fig2}(top) graphs a set of pairs ($v$, $w$) for which $\Delta\beta=\beta(\lambda_{SH})-2\beta(\lambda_{FF})=0$, 
i.e., type I phase matching. For each pair we computed the modal overlap integral and the quadratic susceptibility 
$\Gamma=\iint_{S}d(x,y)(E_{y,FF})^2E_{x,SH}dxdy$, with $S$ the cross section and the fields normalized 
as to yield the  modal power $P=\iint|E(x,y)|^2dxdy$. 
$d(x,y)$ is the spatial distribution of the susceptibility coupling a $\hat{y}$-polarized FF to an $\hat{x}$-polarized SH through the element $d_{14}$. 
In our waveguide, $d(x,y)=d_{14}$ within the $w \times h$ rods - cut on a (001) substrate after a 45° rotation with respect to crystal axes \cite{Berger_JQE} - and $d(x,y)=0$ elsewhere.
As an increase in $w$ forces the mode to fill more the nonlinear dielectric,
but the higher $w$ the higher is the required $v$ to achieve PM, $\Gamma$ exhibits a maximum. 
The corresponding parameters yield the highest conversion efficiency per given propagation length.

\begin{figure}[htb]
\centerline{\includegraphics[width=8.3cm]{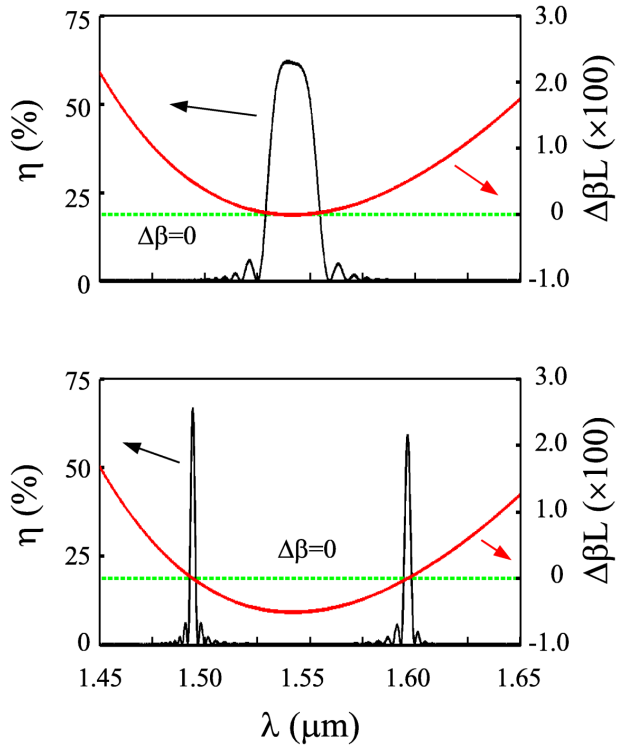}}
 \caption{SHG conversion efficiency (left vertical axes) and phase mismatch (right vertical axes) versus $\lambda_{FF}$ in a 1mm-long structure. Top: optimum $v$ and $w$ from Fig. 3. Bottom: $v$ was reduced by $\sim5\%$ with respect to the top panel. Crossings between the horizontal line and the detuning curve correspond to PM.}
\label{fig3}
\end{figure}

The dependence of SHG efficiency on $\lambda_{FF}$ is graphed in Fig. \ref{fig3} along with the phase mismatch for a slot waveguide of unity length $L=1mm$, accounting for dispersion. Suspended membranes of lengths and aspect-ratios comparable to the one proposed here have been recently realized.\cite{Combrie,Kotlyar} The top panel shows conversion $\eta$ and $\Delta\beta L$ versus $\lambda_{FF}$ for the optimum parameters encircled in Fig. 3 and unity ($1W$) excitation, corresponding to a (material) power density of $\approx 400 MW/cm^2$ and comparable to what previously employed for SHG in GaAs/AlGaAs guides \cite{Moutzouris}. Clearly, a large SHG is associated with the peak of a sinc-like curve and a specific PM  condition (horizontal dashes) near $\lambda_{FF}=1.55\mu m$. Such an individual zero for $\Delta\beta$, however, is not the norm. As displayed in the bottom of Fig. 3, a slight detuning from the optimum $v$ shifts and splits the PM-wavelength, giving rise to a double SH resonance with a mismatch remaining moderate between the two zeroes.  This stems from the strong birefringence of the structure, yielding different dispersion for the two cross-polarized harmonics. For a given void, while TM-like modes substantially undergo material dispersion, TE SH is quite sensitive to wavelength changes, resulting in a multi-peaked SHG conversion efficiency amenable to engineering towards broad-band doublers.

In conclusion, we have proposed and numerically investigated a simple structure for easily tunable and broad-band SHG. The scheme, suitable of straightforward extension to other parametric interactions, permits phase matching even in isotropic crystals, while preserving the nonlinear material response and a large overlap integral. The waveguide with a nano-void, well realizable with current technologies, allows extra tuning by, e.g., the infiltration of thermo- or electro-optic dielectrics or the use of microtransducers for real-time adjustment of the air-gap as well as direct modulation of the fields through the introduction of controlled mismatch (i.e., via cascading \cite{assanto}). We anticipate a whole new family of ultra-compact doublers and mixers.

Acknowledgements. We thank the European Union (2004-512450 POISE) and the Italian Ministry for Research (PRIN 2005098337).

%\pagebreak


\begin{thebibliography}{99}
\bibitem{Franken} P. A. Franken, A. E. Hill, C. W. Peters and G. Weinreich, Phys. Rev. Lett. {\bf 7}, 118 (1961).
\bibitem{Armstrong} J. A. Armstrong, N. Bloembergen, J. Ducuing and P. S. Pershan, Phys. Rev. {\bf 127}, 1918 (1962).
\bibitem{Boyd} R. W. Boyd, \textit{Nonlinear Optics} (Academic Press, New York 2002)
\bibitem{Fejer} M. M. Fejer, G. A.  Magel, D. H. Jundt, R. L. Byer, IEEE J. Quantum Electron. {\bf 28}, 2631 (1992).
\bibitem{Ding} Y. J. Ding, J. B. Khurgin, Opt. Lett. {\bf 21}, 1445 (1996)
\bibitem{Vella_APL1981} P. J. Vella, R. Normandin and G. I. Stegeman, Appl. Phys. Lett. {\bf 38}, 759 (1981)
\bibitem{RavaroJAP_05} M. Ravaro, Y. Seurin, S. Ducci, G. Leo, V. Berger, A. DeRossi, G. Assanto, J. Appl. Phys. {\bf 98}, 063103 (2005)
\bibitem{Baudrier} M. B. Raybaut, R. Haïdar, Ph. Kupecek, Ph. Lemasson, E. Rosencher, Nature {\bf 432}, 374 (2004).
\bibitem{Levenson} C. D'Aguanno, M. Centini, M. Scalora, C. Sibilia, Y. Dumeige, P. Vidakovic, J. A. Levenson, M. J. Bloemer, C. M. Bowden, J. W. Haus, M. Bertolotti, Phys. Rev. E {\bf 64}, 016609 (2001)
\bibitem{DiFalco} A. DiFalco, C. Conti, G. Assanto, Opt. Lett. {\bf 31}, 250 (2006)
\bibitem{VanderZiel} J. P. Van der Ziel, Appl. Phys. Lett. {\bf 26}, 60 (1975)
\bibitem{Fiore} A. Fiore, V. Berger, E. Rosencher, P. Bravetti, J. Nagle, Nature 391, 463 (1998)
\bibitem{Moutzouris} K. Moutzouris, S. V. Rao, M. Ebrahimzadeh,  A. DeRossi, M. Calligaro, V. Ortiz, V. Berger, Appl. Phys. Lett. {\bf 83}, 620 (2003)
\bibitem{Berger_JQE} A. DeRossi, V. Berger, G. Leo and G. Assanto, IEEE J. Quantum Electron. {\bf 41}, 1293 (2005)
\bibitem{Joannopoulos} J. D. Joannopoulos, P. R. Villeneuve, S. Fan, \textit{Photonic Crystals} (Princeton Univ. Press, 1995)
\bibitem{Lipson} Q. Xu, V. R. Almeida, R. R. Panepucci, and M. Lipson, Opt. Lett. {\bf 29}, 1626 (2004)
\bibitem{Afromovitz} M. A. Afromovitz, Solid State Commun. \textbf{15,} 59 (1974)
\bibitem{Combrie} S. Combri{\'e}, A. De Rossi, L. Morvan, S. Tonda, S. Cassette, D. Dolfi and A. Talneau, Electron. Lett. {\bf 42}, 86 (2006)
\bibitem{Kotlyar} M. V. Kotlyar, L. O'Faolain, R. Wilson, T. F. Krauss, J. Vac. Sci. Technol. B {\bf 22}, 1788 (2004).
\bibitem{assanto} G. Assanto, G. Stegeman, M. Sheik-Bahae, E. VanStryland, IEEE J. Quantum Electron. {\bf 31}, 673 (1995).


\end{thebibliography}
\end{document}